\documentclass[aps,preprint,amsmath,amssymb,superscriptaddress,nofootinbib]{revtex4}

\usepackage{graphicx}
\begin{document}

\title{Semi-elastic cross section for a scalar resonance of mass 750 GeV}

\author{\.{I}. \c{S}ahin}
\email[]{inancsahin@ankara.edu.tr}
 \affiliation{Department of
Physics, Faculty of Sciences, Ankara University, 06100 Tandogan,
Ankara, Turkey}

\begin{abstract}
We assume that the recently reported excess over Standard Model
expectations in the diphoton production is originated from a new
scalar boson $\phi$ which interacts with photons through the
coupling $\propto \phi F^2$. We obtain the cross section for
semi-elastic production process $pp\to p \gamma p\to p \phi q X$ by
considering $\phi \gamma \gamma$ and $\phi \gamma Z$ couplings and
discuss the constraints on the coupling parameters. We investigate
the potential of the semi-elastic production process to probe $\phi
\gamma \gamma$ and $\phi \gamma Z$.

\end{abstract}


\maketitle

\section{Introduction}
Recently a new resonance of mass around 750 GeV has been reported by
the ATLAS and CMS Collaborations \cite{ATLAS1,CMS1}. Although it is
not confirmed with a significant statistics, we may expect that it
is a genuine signal coming from new physics beyond the Standard
Model. We assume that the signal is real and caused by a new neutral
scalar boson $\phi$ which interacts with gauge bosons. We will
ignore its coupling to gluons and quarks since no evidence for
resonant particles is observed up to now in the dijet final state
\cite{Khachatryan:2015dcf}. The new scalar resonance can be
investigated in a model-independent way by means of the effective
Lagrangian formalism. In writing effective interactions of $\phi$
with gauge bosons we employ the formalism presented in
Refs.\cite{Fichet:2015yia,Fichet:2015vvy}. The effective Lagrangian
which describes $\phi \gamma \gamma$ and $\phi \gamma Z$ couplings
is given by
\begin{eqnarray}
\label{effectivelagrangian} {\cal L}=c_{\gamma \gamma} \phi
F^{\mu\nu}F_{\mu\nu}+c_{\gamma z} \phi F^{\mu\nu} Z_{\mu\nu}
\end{eqnarray}
where $F_{\mu\nu}=\partial_\mu A_\nu-\partial_\nu A_\mu$ and
$Z_{\mu\nu}=\partial_\mu Z_\nu-\partial_\nu Z_\mu$. Here, $A$ and
$Z$ are photon and Z boson fields. The coupling constants $c_{\gamma
\gamma}$ and $c_{\gamma z}$ have the dimension of inverse energy.
They are related to the coefficients of the effective operators
before the symmetry breaking through $c_{\gamma
\gamma}=\cos^2{\theta_W}f_B^{-1}+\sin^2{\theta_W}f_W^{-1}$ and
$c_{\gamma z}=2\sin{\theta_W}
\cos{\theta_W}\left(f_W^{-1}-f_B^{-1}\right)$
\cite{Fichet:2015yia,Fichet:2015vvy}.

The origin of this new resonance has been discussed in a vast number
of studies in the literature. These studies cover a wide range of
new physics scenarios such as string inspired models, gauge symmetry
models, radiative and top seesaw models, Next-to-Minimal
Supersymmetric Standard Model, model independent analysis via
effective lagrangian,
etc.\cite{Fichet:2015vvy,Csaki:2015vek,Anchordoqui:2015jxc,Fichet:2016pvq,scenarios}.
The production of the scalar resonance has been investigated in
elastic proton-proton scattering in
Refs.\cite{Fichet:2015vvy,Csaki:2015vek,Anchordoqui:2015jxc,Fichet:2016pvq}.
In Refs.\cite{Fichet:2015vvy,Fichet:2016pvq} authors also considered
inelastic photon-photon fusion and took into account of the
effective couplings $\phi\gamma\gamma, \phi ZZ,\phi Z\gamma$ and
$\phi WW$. In this paper we will consider the semi-elastic
production of the scalar boson $\phi$ via the process $pp\to p
\gamma p\to p \phi q X$. Some part of the cross section in the
photon-photon fusion should belong to semi-elastic processes. It is
therefore important to know the size of the cross section for
semi-elastic production of the scalar boson. Furthermore, with the
aid of forward detectors semi-elastic processes can be efficiently
discerned from inelastic processes. They can be considered in
isolation from inelastic processes and can be used to perform
precision measurements. Therefore it is important to discover the
potential of the semi-elastic production processes to probe $\phi$
couplings to Standard Model fields.

\section{Cross sections and numerical results}
\begin{figure}
\includegraphics[scale=0.7]{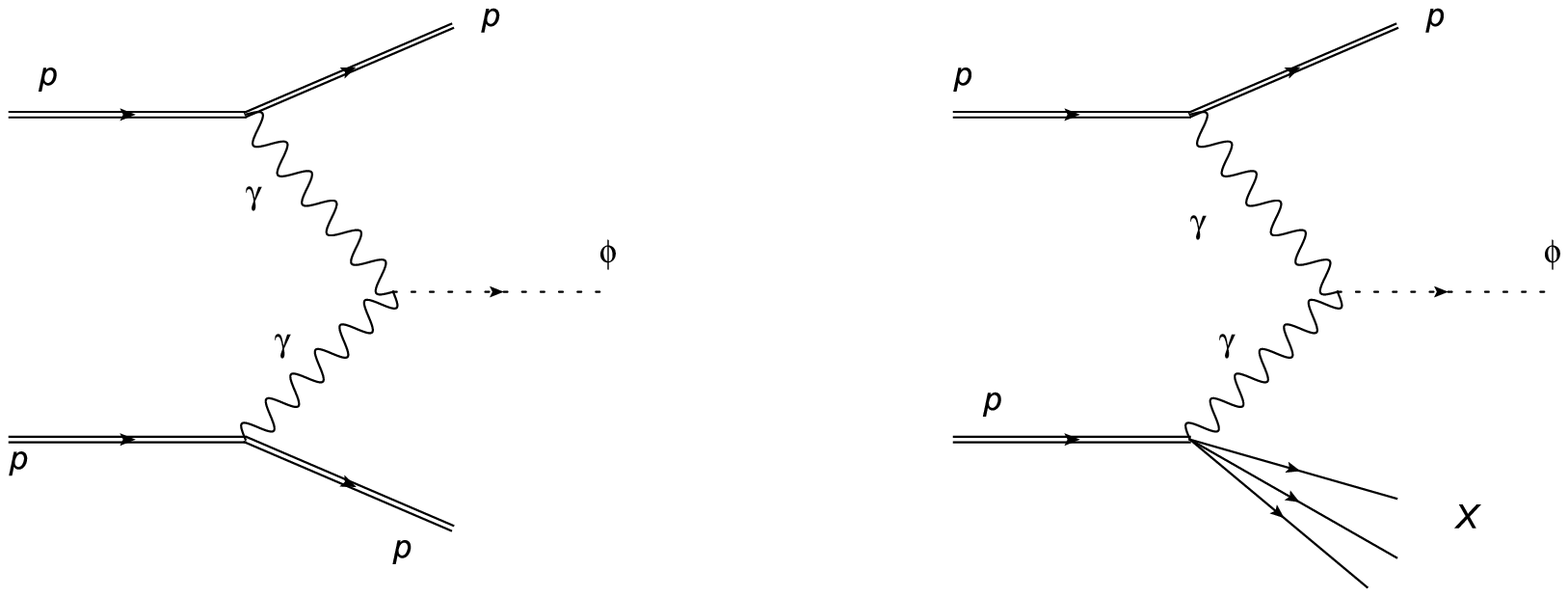}
\caption{ Schematic diagram for elastic (left panel) and semi-elastic (right panel)
production of the scalar boson in a photon-photon fusion.
\label{fig1}}
\end{figure}

In an elastic proton-proton scattering incoming protons do not
dissociate into partons but they remains intact.
On the other hand, in a semi-elastic proton-proton scattering,
one of the incoming proton dissociates into partons but
the other proton remains intact \cite{rouby,Schul:2011xwa}.
Fig.\ref{fig1} represents elastic and semi-elastic production of
the scalar boson in a photon-photon fusion. The elastic photon emission from
the proton can be described by equivalent photon approximation.
In equivalent photon approximation we employ the formalism presented in Refs.\cite{budnev1975,baur2002,piotrzkowski2001}
where the electromagnetic form factors of the proton are taken into consideration.

\begin{figure}
\includegraphics[scale=0.7]{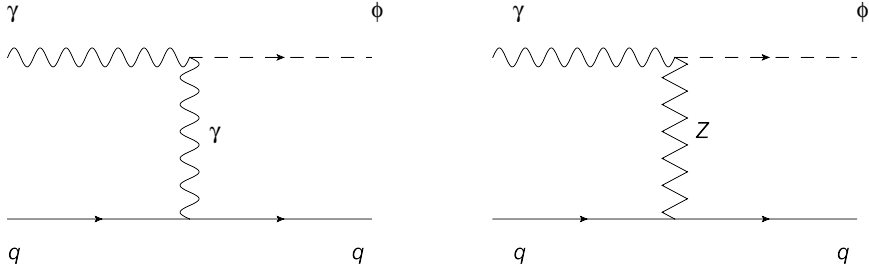}
\caption{Tree-level Feynman diagrams for the subprocess $\gamma q\to \phi q$.
\label{fig2}}
\end{figure}

The process $pp\to p \gamma p\to p \phi q X$ comprise of the subprocesses $\gamma q\to \phi q$ where
$q$ can be $u,d,s,c,b$ quarks or anti-quarks. This gives totally 10 independent subprocess. Each of the subprocesses
is described by t-channel $\gamma$ and $Z$ exchange diagrams (Fig.\ref{fig2}). Hence both photon-photon
and photon-Z boson fusion contribute to the $\phi$ production. The cross section for semi-elastic production
$pp\to p \gamma p\to p \phi q X$ can be obtained through the integration
\begin{eqnarray}
\label{mainprocess}
 \sigma\left(pp\to p \gamma p\to p \phi q X\right)=\sum_q\int_{\xi_{min}}^{\xi_{max}} {dx_1 }\int_{0}^{1}
{dx_2}\left(\frac{dN_\gamma}{dx_1}\right)\left(\frac{dN_q}{dx_2}\right)
\hat{\sigma}_{\gamma q\to \phi q}(\hat s).
\end{eqnarray}
Here, $\frac{dN_\gamma}{dx_1}$ is the equivalent photon spectrum.
Its analytical expression was defined in
Refs.\cite{budnev1975,baur2002,piotrzkowski2001,kepka2008}.
$\frac{dN_q}{dx_2}$ is the quark distribution function of the
proton. We evaluate it numerically by using a code MSTW2008
\cite{Martin:2009iq}. $x_1$ represents the fraction $E_\gamma/E$
where $E_\gamma$ and $E$ are the energy of the emitted photon and
the proton. $x_2$ is the momentum fraction of the proton's momentum
carried by the quark.  The limits of the $dx_1$ integration is
determined by the upper and lower bounds of the parameter $\xi$
which represents the momentum fraction loss of the proton. After
elastic photon emission protons generally deviate slightly from the
direction of beam pipe and escape from the central detectors without
interacting. Forward detectors can detect these intact scattered
protons. The acceptance of the forward detectors is determined by
the interval $\xi_{min}<\xi<\xi_{max}$. The use of forward detectors
is especially necessary for high-luminosity runs at the LHC in order
to isolate elastic or semi-elastic processes
\cite{Albrow:2008pn,Tasevsky:2009zza,Albrow:2010yb,Tasevsky:2014cpa}.
We will consider two different cases: In the first case we do not
assume the existence of the forward detectors and take account of
the whole interval $0<x_1<1-\frac{m_p}{E}$ for the $dx_1$
integration. This case makes sense if we want to examine the size of
the cross section coming from semi-elastic processes to the whole
production but do not intend to isolate the semi-elastic production.
In the second case we consider a forward detector acceptance of
$0.015<\xi<0.15$
\cite{Fichet:2014uka,forward-new1,forward-new2,Tasevsky:2015xya}.

\begin{figure}
\includegraphics[scale=1]{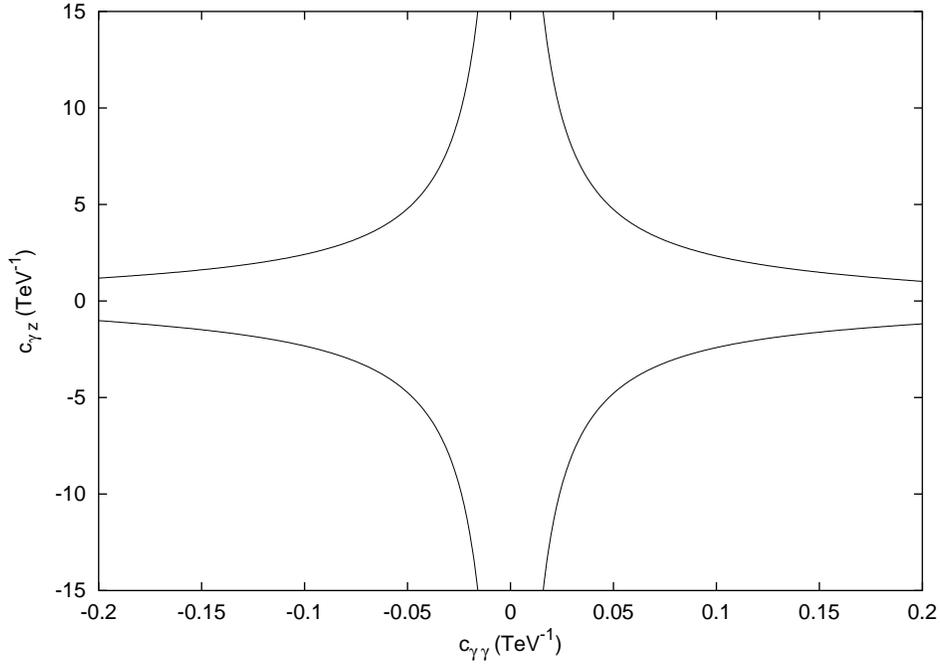}
\caption{The area restricted by the lines corresponds to the values
of the total cross section $\sigma\left(pp\to p \gamma p\to p \phi q
X\right)BR\left(\phi\to \gamma \gamma\right)\leq5\;fb$.\label{fig3}
We sum all contributions from subprocesses $\gamma q\to \phi q$ for
$q=u,d,s,c,b,\bar u,\bar d,\bar s,\bar c,\bar b$. The center-of-mass
energy of the proton-proton system is taken to be $\sqrt s=13\;
TeV$. We consider the whole interval $0<x_1<1-\frac{m_p}{E}$.}
\end{figure}

\begin{figure}
\includegraphics[scale=1]{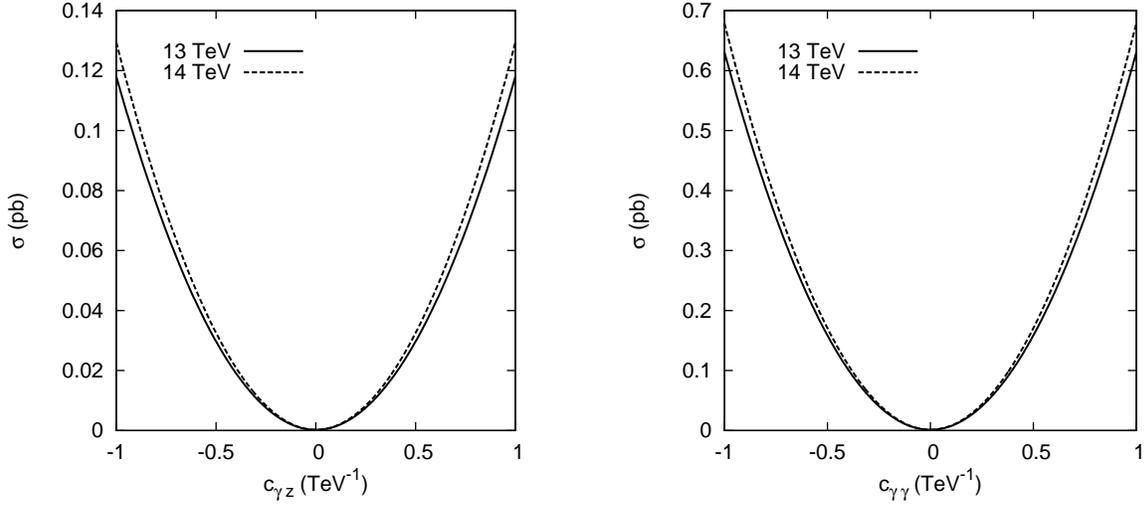}
\caption{Total cross section of $pp\to p \gamma p\to p \phi q X$ as
a function of the coupling $c_{\gamma z}$ (left panel) and
$c_{\gamma \gamma}$ (right panel) for two different LHC
center-of-mass energies stated on the figures. We sum all
contributions from subprocesses $\gamma q\to \phi q$ for
$q=u,d,s,c,b,\bar u,\bar d,\bar s,\bar c,\bar b$ and consider the
whole interval $0<x_1<1-\frac{m_p}{E}$. Each time only one of the
coupling parameters have been kept different from zero.
\label{fig4}}
\end{figure}

\begin{figure}
\includegraphics[scale=1]{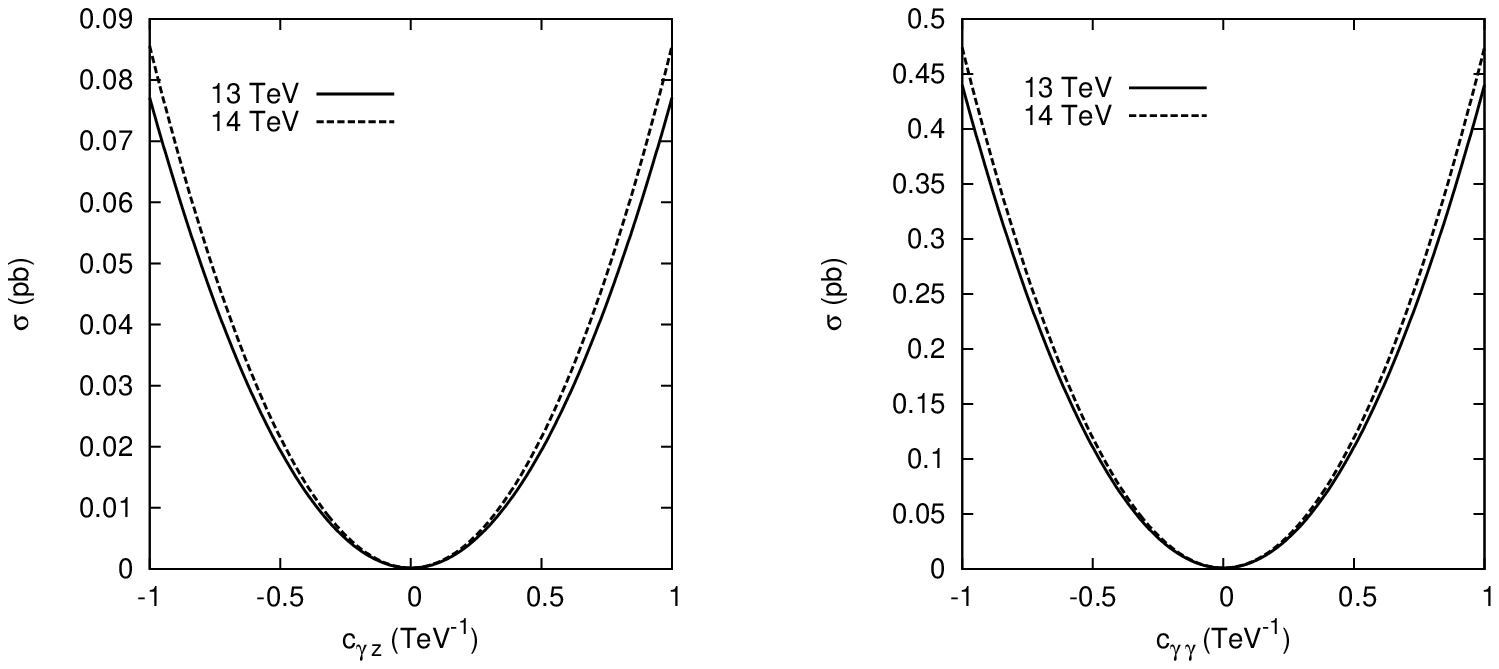}
\caption{Total cross section of $pp\to p \gamma p\to p \phi q X$ as
a function of the coupling $c_{\gamma z}$ (left panel) and
$c_{\gamma \gamma}$ (right panel) for two different LHC
center-of-mass energies stated on the figures. We sum all
contributions from subprocesses $\gamma q\to \phi q$ for
$q=u,d,s,c,b,\bar u,\bar d,\bar s,\bar c,\bar b$ and consider the
forward detector acceptance of $0.015<\xi<0.15$. Each time only one
of the coupling parameters have been kept different from zero.
\label{fig5}}
\end{figure}

Experimental results indicate that this new resonance has a total
width of $\Gamma_{total}\approx 45\;GeV$ and the observed number of
events corresponds to $\sigma(pp\to \phi X)BR(\phi\to \gamma
\gamma)\approx3-6\;fb$ \cite{ATLAS1,CMS1}. Here the branching ratio
is given by $BR(\phi\to \gamma \gamma)=\frac{c_{\gamma\gamma}^2
m_\phi^3}{4\pi\Gamma_{total}}$. To give an idea about size of the
cross section coming from semi-elastic production we solve the
equation $\sigma(pp\to p \gamma p\to p \phi q X)BR(\phi\to \gamma
\gamma)=5\;fb$ numerically for the couplings $c_{\gamma\gamma}$ and
$c_{\gamma z}$. Corresponding plots are presented in Fig.\ref{fig3}.
In the figure the area restricted by the lines corresponds to the
values of the total cross section less than $5 fb$. Therefore the
cross section for semi-elastic production of the scalar boson allows
the values of $c_{\gamma\gamma}$ and $c_{\gamma z}$ within this
restricted area. In Figs.\ref{fig4} and \ref{fig5} we plot the total
cross section of $pp\to p \gamma p\to p \phi q X$ as a function of
the couplings $c_{\gamma z}$ and $c_{\gamma \gamma}$ for the LHC
center-of-mass energies of $\sqrt s=13\;TeV$ and $14\;TeV$. In
Fig.\ref{fig4} we consider the whole interval
$0<x_1<1-\frac{m_p}{E}$ for the $dx_1$ integration in
Eq.(\ref{mainprocess}). But in Fig.\ref{fig5} we consider the
forward detector acceptance of $0.015<\xi<0.15$. We observe from
these figures that the cross section for the acceptance of
$0.015<\xi<0.15$ is approximately a factor of $1.5$ smaller than the
cross section for $0<x_1<1-\frac{m_p}{E}$.

\begin{figure}
\includegraphics[scale=1]{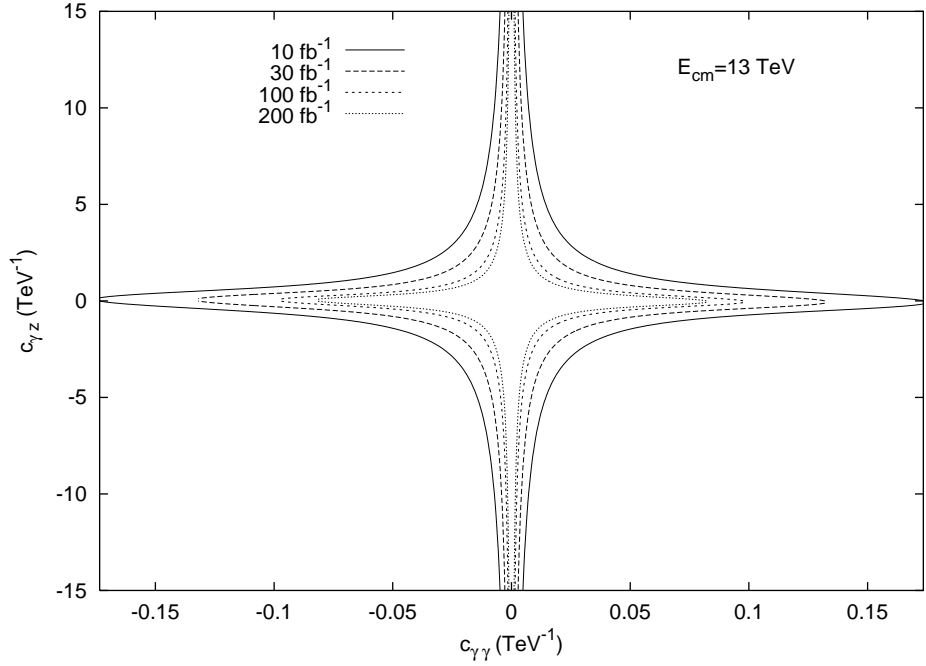}
\caption{The areas restricted by the lines represent $95\%$
confidence level sensitivity bounds on the parameter space
$c_{\gamma \gamma}-c_{\gamma z}$. The legends are for various LHC
luminosities. The center-of-mass energy of the colliding protons is
taken to be $\sqrt s=13\; TeV$. \label{fig6}}
\end{figure}

In order to investigate the potential of the process $pp\to p \gamma
p\to p \phi q X$ to probe $\phi \gamma \gamma$ and $\phi \gamma Z$
couplings in a future experiment we have performed a statistical
analysis using a Poisson distribution. The expected number of events
has been calculated through the formula $N=\sigma(pp\to p \gamma
p\to p \phi q X)BR(\phi\to \gamma \gamma)L_{int}$ where $L_{int}$ is
the integrated luminosity. We assume that the number of observed
events $N_{obs}$ equal to the Standard Model prediction. The
determination of an on-shell scalar boson with mass
$m_\phi\approx750\;GeV$ requires an invariant mass measurement of
the final state photon pair. Therefore, a cut of
$M_{\gamma\gamma}\approx750\;GeV$ should be imposed on the invariant
mass of final state photons. This cut reduces the contribution
coming from background processes. Hence we ignore any background
contribution and assume that $N_{obs}=0$. This gives an upper limit
of $N_{up}=3$ for number of events at $95\%$ confidence level. In
Fig.\ref{fig6} we show $95\%$ confidence level sensitivity bounds on
the parameter space $c_{\gamma \gamma}-c_{\gamma z}$ for integrated
luminosities of $L_{int}=10,\;30,\;100$ and $200\;fb^{-1}$ and
forward detector acceptance of $0.015<\xi<0.15$. The center-of-mass
energy of the colliding protons is taken to be $\sqrt s=13\; TeV$.
We have also calculated the bounds for center-of-mass energy of
$\sqrt s=14\; TeV$. We do not present the bounds for $\sqrt s=14\;
TeV$ since the bounds for $\sqrt s=14\; TeV$ are very close to the
bounds for $\sqrt s=13\; TeV$. The difference between the bounds for
these two different center-of-mass energies does not exceed $2\%$.

\section{Conclusions}

The inelastic production of the scalar boson receive contributions
both from $\gamma\gamma,ZZ,Z\gamma$ and $WW$ fusion due to effective
$\phi\gamma\gamma, \phi ZZ,\phi Z\gamma$ and $\phi WW$ couplings. On
the other hand, the semi-elastic production $pp\to p \gamma p\to p
\phi q X$ receives contributions only from $\gamma\gamma$ and
$\gamma Z$ fusion and give us the opportunity to probe
$\phi\gamma\gamma$ and $\phi Z\gamma$ couplings independent from
$\phi ZZ$ and $\phi WW$. Another advantage of the semi-elastic
production is that it provides a rather clean channel compared to
inelastic production due to the absence of one of the incoming
proton remnants. Moreover, in principle forward detectors can detect
the momentum loss of intact scattered protons. The knowledge
obtained in this way may be useful in reconstructing the kinematics
of the subprocesses. Hence, the semi-elastic production $pp\to p
\gamma p\to p \phi q X$ process can be considered as a candidate for
precision measurements. It is obvious that the production of $\phi$
via elastic proton-proton scattering provides the most clean channel
due to the absence of the remnants of both proton beams. But the
elastic production has a lower energy reach and effective luminosity
with respect to semi-elastic production.

\end{document}